\documentstyle[aps,preprint,floats,psfig]{revtex}
\begin{document}
\newcommand{\beq}{\begin{equation}}
\newcommand{\eeq}{\end{equation}}
\draft
\title  {Resonant flux motion and $I-V$-characteristics in
frustrated Josephson junctions}
\author{ N. Stefanakis
}
\address{CNRS - CRTBT, 25 Avenue des Martyrs
	BP 166-38042 Grenoble c\'edex 9, France}
\date{\today}
\maketitle

\begin{abstract}

We describe the dynamics of fluxons moving in a frustrated Josephson 
junction with $p$, $d$, and $f$-wave symmetry and calculate the $I-V$ 
characteristics. The behavior of fluxons is quite distinct in the 
long and short length junction limit. For long junctions the intrinsic 
flux is bound at the center and the moving integer fluxon or antifluxon
interacts with it only when it approaches the junction's center. 
For small junctions the intrinsic flux can move 
as a bunched type fluxon
introducing additional steps in the $I-V$ 
characteristics. Possible realization in quantum computation is presented. 
 
\end{abstract}
\newpage

\section{introduction}
The determination of the order parameter symmetry in high-$T_c$ 
superconductors is a problem which has not yet been completely solved 
\cite{scalapino,vanh,tsuei,stefan1,stefan2,stefan4}.
The Josephson effect provides a phase sensitive mechanism
to study the pairing symmetry of unconventional superconductors.
In Josephson junctions involving unconventional
superconductors, the sign change of the order parameter
with angle measured from the $x$-axis
in the $ab$ plane introduces an intrinsic phase
shift of $\pi$ in the Josephson current phase relation or alternatively
a negative Josephson critical current.
The effect of shifting the
phase by $\pi$ is equivalent to the shifting the critical current 
versus the magnetic flux pattern
in a squid that contains a $\pi$ junction (called frustrated junction)
by $\Phi_0/2$, where $\Phi_0=h/2e$ is the flux quantum \cite{vanh}. 

The presence of spontaneous or trapped flux is a general property of 
systems where a sign change of the pair potential occurs in orthogonal 
directions in $k$-space. Its existence has been predicted 
for example in ruthenates \cite{maeno} where 
the pairing state is triplet as indicated in the 
Knight shift measurements \cite{ishida}
and the time-reversal symmetry is broken 
as shown by the muon spin rotation ($\mu SR$) experiment \cite{luke}
where the evolution of the polarization of the implanted muon in the 
local magnetic environment of the superconductor gives information 
about the presence of spontaneous magnetic field.   
Moreover the pairing state 
has line nodes within the
gap as indicated by the specific heat measurements \cite{nishizaki}. 
This spontaneous flux shows a characteristic 
modulation with the misorientation angle within the RuO$_2$-plane 
that can be checked by experiment \cite{stefan3}.

The one dimensional Josephson junction 
with total reflection at the end boundaries, 
between $s$-wave superconductors, supports
modes of resonant
propagation of fluxons\cite{fulton}. In the plot of the
current-voltage ($I-V$) characteristics these modes appear as
near-constant voltage branches known as zero field steps (ZFS)
\cite{chen,lomdahl,vernik}.
They occur in the absence of any external field.
The ZFS appear at integer multiple of $V_1=\Phi_0 c_S/l$, where
$c_S$ is the velocity of the 
electromagnetic waves in the junction, and $l$ is the junction
length.
The moving fluxon
is accompanied by a voltage pulse which can be detected at the
junction's edges.

When the contact between a $0$ and $\pi$ junctions, that contains an
intrinsic half-fluxon, is current-biased
the half-fluxon becomes unstable for certain values of the external current
with respect to transforming into an anti-half-fluxon  
and emitting 
an integer fluxon \cite{kuklov}.
Also when a $0-\pi-0$ junction, that contains two half vortices,
is current-biased, for certain critical current, a transition occurs 
between the two degenerate fluxon configurations 
and a voltage pulse is generated \cite{kato}.

In this paper we study the dynamic properties of fluxons and 
calculate the $I-V$ characteristics
in frustrated junctions 
with $B_{1g}$, $E_u$, $B_{1g}\times E_u$ pairing symmetry. 
The last two are candidates pairing states for ruthenates \cite{maeno}. 
The nodeless $p$-wave order parameter
with $E_u$ symmetry has been proposed by Rice and Sigrist \cite{rice} while the 
$B_{1g}\times E_u$ has been proposed by Hasegawa {\it et al.} \cite{hasegawa}.
In junctions involving unconventional superconductors 
the
behavior of fluxons is typically different 
in the long and short length 
junction limit. In the long limit the fractional fluxon is confined at the 
center and the moving fluxon interacts with it only when it approaches the 
center. However in the short limit the bound fluxon becomes able to 
move as a bunched type 
solution with integer or half integer magnetic flux.
For the $B_{1g}$ case the $I-V$ pattern is shifted by a voltage 
that corresponds to the intrinsic phase shift.
Also the frustrated Josephson junction can be considered as a way to build
a quantum 'bit' (qubit) which is 
the generalization of the 'bit' of the classical
computer.

The article is organized as follows. In Sec. II we
develop the model and discuss the formalism. 
In Sec. III we
present the results for the
long junction and in  Sec. IV
for the shorter junction limit.
In Sec. V we discuss the implementation of the qubit
and finish with the conclusions.

\section{corner junction model}
\begin{figure}
  \centerline{\psfig{figure=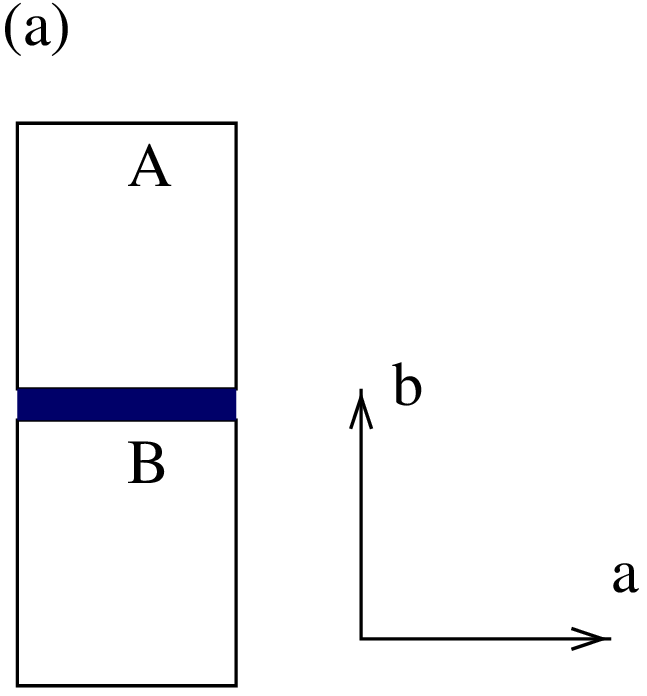,width=5.5cm,angle=-90}}
  \centerline{\psfig{figure=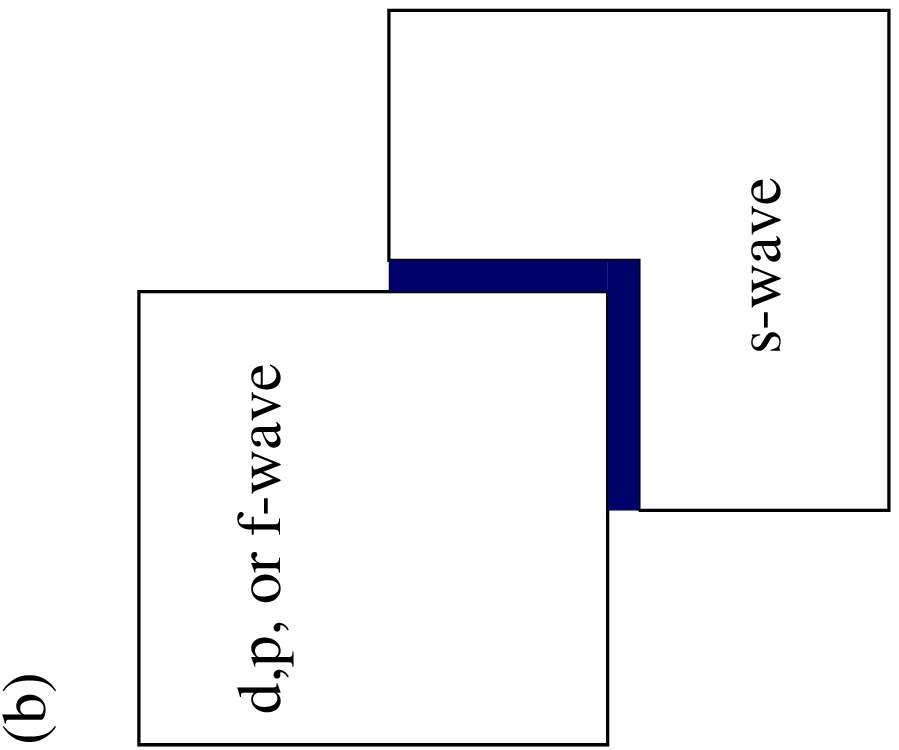,width=5.5cm,angle=-90}}
\caption{
(a) A single Josephson junction between superconductors $A$ and
$B$ with a two component order parameter. Also a small
co-ordinate system indicating $a$ and $b$ crystalline directions 
is shown.
(b) The geometry of the corner junction between
a mixed symmetry superconductor and an $s$-wave superconductor.}
  \label{fig1.fig}
\end{figure}

We consider the junction shown in Fig. 1(a) between a superconductor $A$
with a two component order parameter and a superconductor $B$
with $s$-wave symmetry.
The supercurrent density can be written as:
\begin{equation}
J(\phi)=\widetilde J_c \sin(\phi+ \phi_c),~~~\label{jphidis}
\end{equation}
where $\widetilde J_c$ is the Josephson critical current density,  
$\phi$ is the relative phase difference between the two superconductors and  
$\phi_c$ is the intrinsic phase shift. 
We describe a frustrated junction of length $l$, i.e., 
the two segments have different characteristic phases, i.e., 
$\phi_{c1}$ in $0<x<\frac{l}{2}$ and
$\phi_{c2}$ in $\frac{l}{2}<x<l$.
By introducing an extra relative phase in one part of this junction, 
this one dimensional junction can be mapped in the corner junction
that is seen in Fig. \ref{fig1.fig} (b). 
The characteristic phases $\phi_{c1}$ and $\phi_{c2}$ distinguish
the various pairing symmetries 
and can be seen 
in table \ref{tablephic}.
For the orientation of the junction that we consider in which 
the $a$ and $b$ 
crystal axes are at right angles to the interface a simple 
calculation \cite{stefan1,stefan2,stefan3} gives
$\widetilde J_c=1$
for the pairing states that we consider.

The phase difference across the junction is
then the solution of the time dependent sine-Gordon equation
\begin{equation}
\frac{d^2 { \phi}}{dx^2} - \frac{d^2 { \phi}}{dt^2}  =
J(\phi)+\gamma \frac{d{ \phi}}{dt}
,~~~\label{eq01}
\end{equation}
with the following inline boundary conditions 
\begin{equation}
\frac{d { \phi}}{dx}\left|_{x=0,l}\right. =\pm \frac{{
I}}{2},
~~~\label{eq02}
\end{equation}
where 
the time $t$ is in units $\omega_0^{-1}$, where
$\omega_0=\sqrt{\frac{2 \pi \widetilde J_c}{\Phi_0 C}}$ is 
the Josephson plasma frequency, $C$ is the 
capacitance per unit length.
$\gamma=G/\omega_0 C$ is the damping constant which depends on the temperature, 
$G^{-1}$ is an effective normal resistance. 
The value used in the numerical calculations is $\gamma=0.01$.
The length $x$ is normalized in units of the
Josephson penetration depth 
$\lambda_J=\sqrt{\frac{\Phi_0}{2\pi \widetilde J_c L_p}}$, 
$L_p$ is the inductance per length and is given by 
$L_p=\mu_0 d$, where $d=2\lambda_L + t_{ox}$ is 
the magnetic thickness of the junction, $\lambda_L$ is the 
London penetration depth, $t_{ox}$ is the thickness 
of the insulating oxide layer, and $\mu_0=4\pi 10^{-7} H/m$.
The velocity of the
electromagnetic waves in the junction is given by 
$c_S=\sqrt{L_p C}$.
$I$ is the normalized inline bias current in units of 
$\lambda_J \widetilde J_c$.

In previous publication \cite{stefan2} 
we used overlap boundary conditions, where 
the current is uniformly distributed in space. 
However in actual experiments in the $s$-wave case, 
the biased current in the 
overlap geometry may be concentrated at the edges within a 
length $\lambda_J$ rather than distributed in space \cite{scheuermann}. 
Therefore it 
is more appropriate to use inline boundary conditions. 
However in the case of the overlap geometry 
only details of the fluxon propagation are 
quantitatively different e.g. 
the oscillations of the bound 
fluxons about their equilibrium position 
and their interaction with the moving fluxons.
However the basic physics of the problem i.e. the shift of the 
voltage values is independent on the choice of the boundary conditions. 

\section{long junction limit}
\begin{figure}
\centerline{\psfig{figure=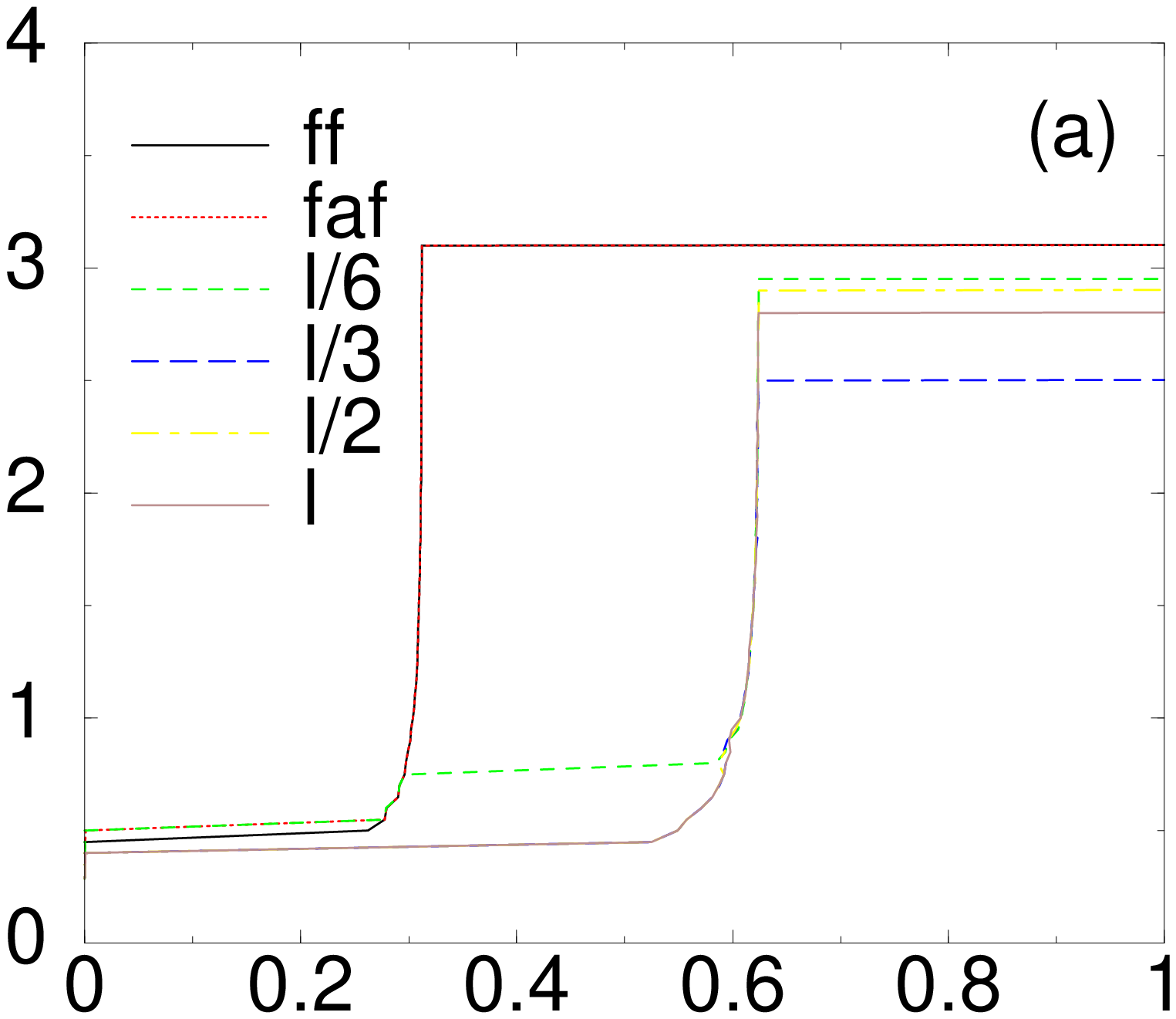,width=5.5cm,angle=0}}
\centerline{\psfig{figure=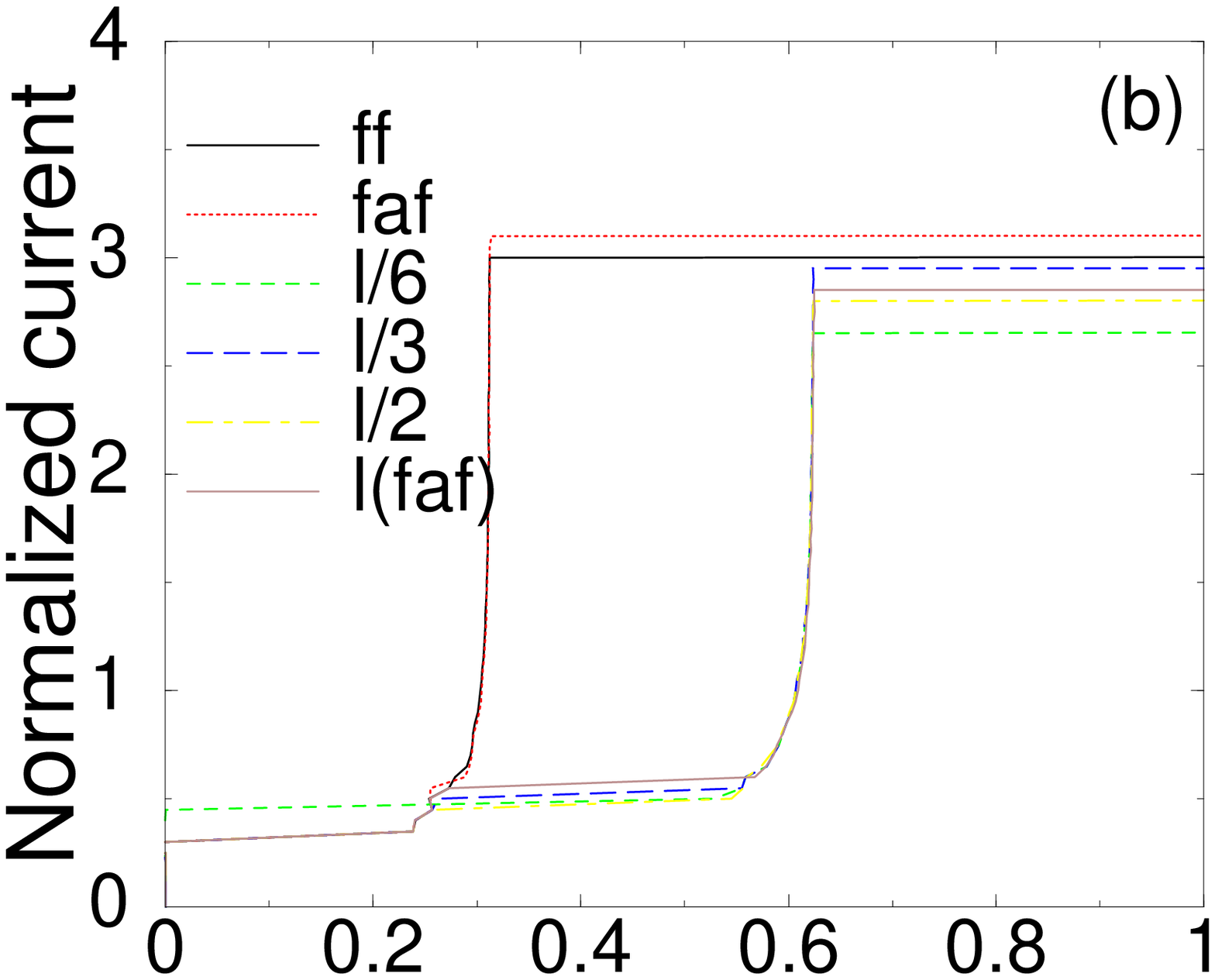,width=5.5cm,angle=0}}
\centerline{\psfig{figure=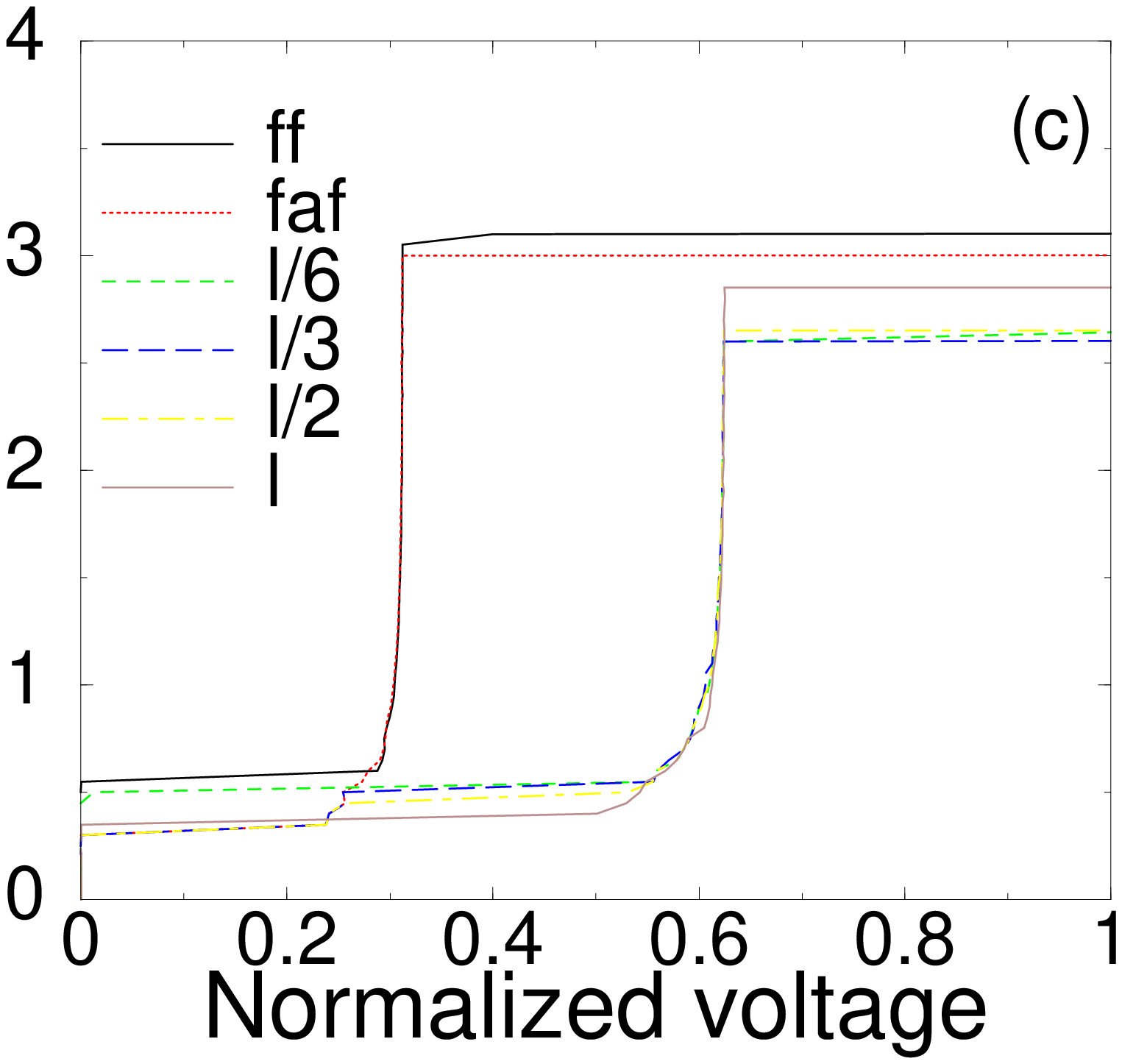,width=5.5cm,angle=0}}
\caption{Normalized current versus the normalized voltage 
for the inline geometry, for the first and 
second ZFS. The solutions for the first ZFS are the $ff$, $faf$ corresponding 
to a bound fluxon or antifluxon in the junction's center.
For the second ZFS the solutions are labeled by the relative distance 
between the fluxons $l/x$, 
where $l=20$ is the junction length,
and $x=1,2,3,6$ respectively, $\gamma=0.01$.
The pairing state is (a) $B_{1g}$-wave, (b) $E_u$-wave, 
(c) $B_{1g}\times E_u$-wave.
}
\label{IV.fig}
\end{figure}

\begin{figure}
\centerline{\psfig{figure=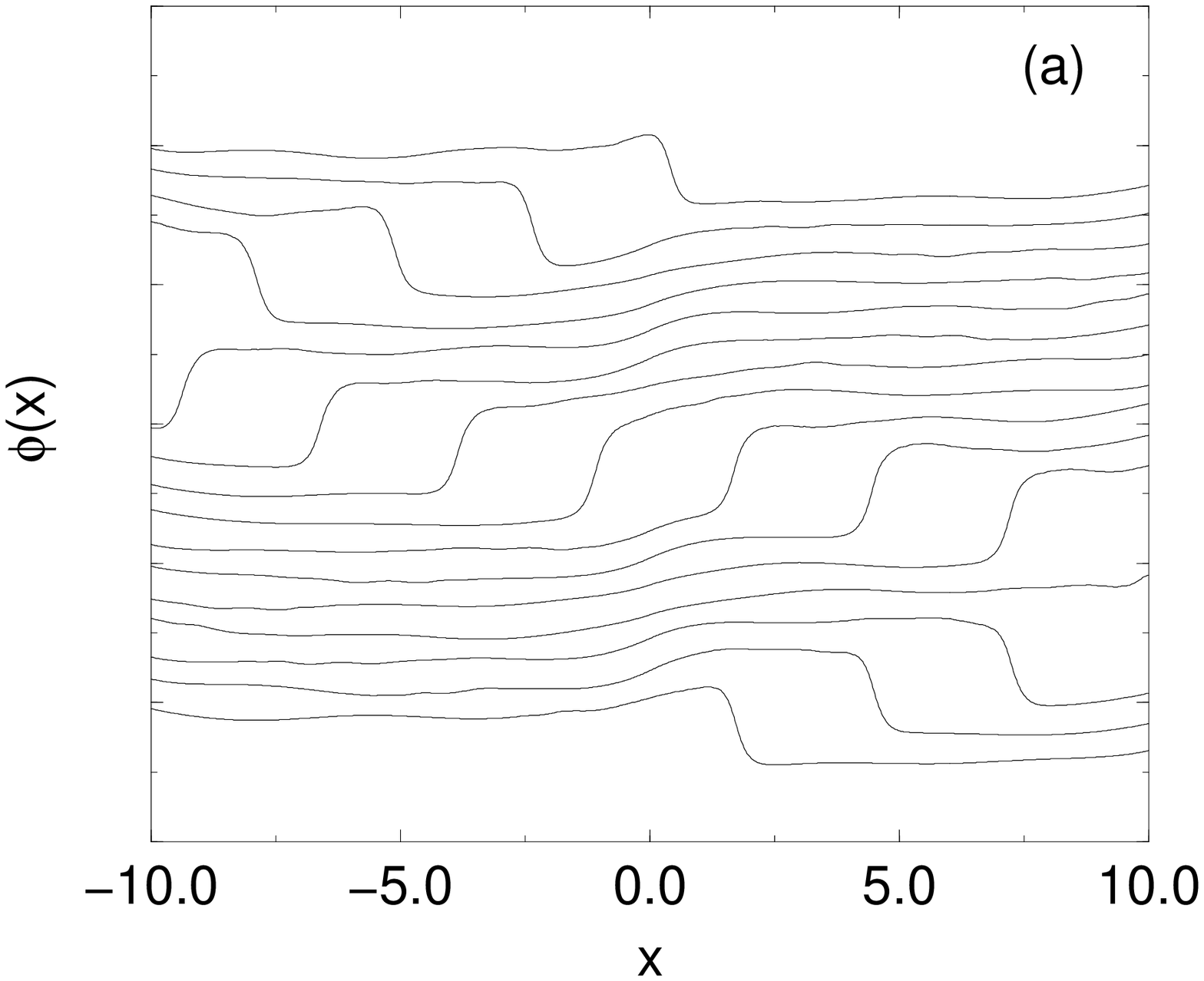,width=8.5cm,angle=0}}
\centerline{\psfig{figure=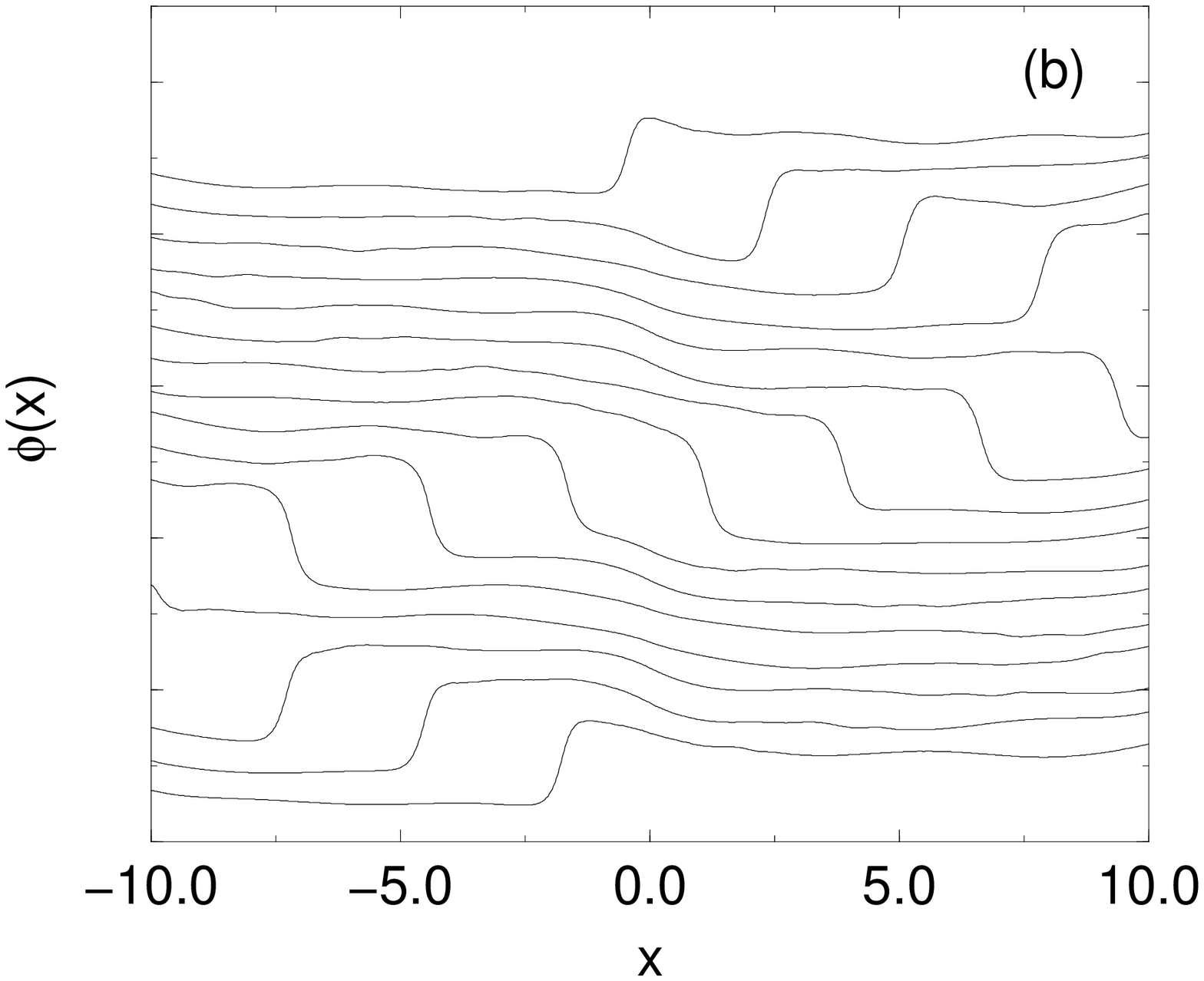,width=8.5cm,angle=0}}
\caption{Phase $\phi(x)$ vs $x$
for the solutions in the first ZFS,
at various instants, during one period
separated by $\Delta \tau=2.8$.
The curves are shifted by $0.5$ to avoid overlapping.
$l=20$, $I=1.6$, $\gamma=0.01$: (a) $ff$,
(b) $faf$. The pairing state is $B_{1g}$-wave.} 
\label{1ZFS.fig}
\end{figure}

A $4^{th}$ order Runge Kutta method with fixed time step $\Delta t=0.01$, 
was used for 
the integration of the equations of motion. The number of grid points is $N=1000$.
We discuss first the case where the junction length is long
$l=20$. We present in Fig. \ref{IV.fig} 
the $I-V$
characteristics for the first and 
second ZFS that correspond to the case where one or two fluxons are 
moving into the junction. The pairing state is $B_{1g}$ (a), $E_u$ (b), 
$B_{1g}\times E_u$ (c).
For the $B_{1g}$ the external current cannot move the fractional fluxon ($ff$)
which is confined at $x=0$ (see Fig. \ref{1ZFS.fig}(b)). 
However for certain value of the bias current the $ff$ is transformed into 
an fractional antifluxon ($faf$) and an integer fluxon ($F$) 
is emitted which is traveling to the left. 
The $F$ hits the left boundary and transforms
into an integer antifluxon ($AF$) which moves to the right. 
When the $AF$ reaches the center
interacts with the $faf$ but is not able to change its polarity and results 
into an $faf$ and an $AF$ moving to the right.
The antifluxon hits the right boundary transforms into a fluxon
which moves to the center where it meets the oscillating $faf$ and interacts
with it forming a $ff$ and the period is completed.

When a $faf$ exists at the junction center, by applying the external
current it emits an $AF$ which moves to the right
and it converts itself to a $ff$ (see Fig. \ref{1ZFS.fig}(a)). 
The $AF$ hits the right boundary and transforms
into a $F$ which moves to the left. When the $F$ reaches the center
interacts with the $ff$ and results into a $ff$ and a $F$ moving to the left.
The fluxon hits the left boundary transforms into an antifluxon
which moves to the center where it meets the oscillating $ff$ and interacts
with it forming a $faf$ and the period is completed.

In the relativistic limit $c_S \approx 1$ 
reached at high currents a full period 
of motion back and forth takes time $t=2l/c_S=40$ and since the overall phase 
advance is $4\pi$ the normalized 
voltage will be $V=\phi_t=\frac{4\pi}{40}=0.314$
for the first ZFS. So when the 
junction length is large, the ZFS 
occur at the same values of the dc-voltage 
independently from the pairing symmetry, since one 
full fluxon or antifluxon propagates in the junction.
These values much exactly the ones for 
conventional $s$-wave superconductors junction.
The direction of the voltage pulse depends on the sign of the intrinsic flux 
and can be used for the qubit implementation. 

\begin{figure}
\centerline{\psfig{figure=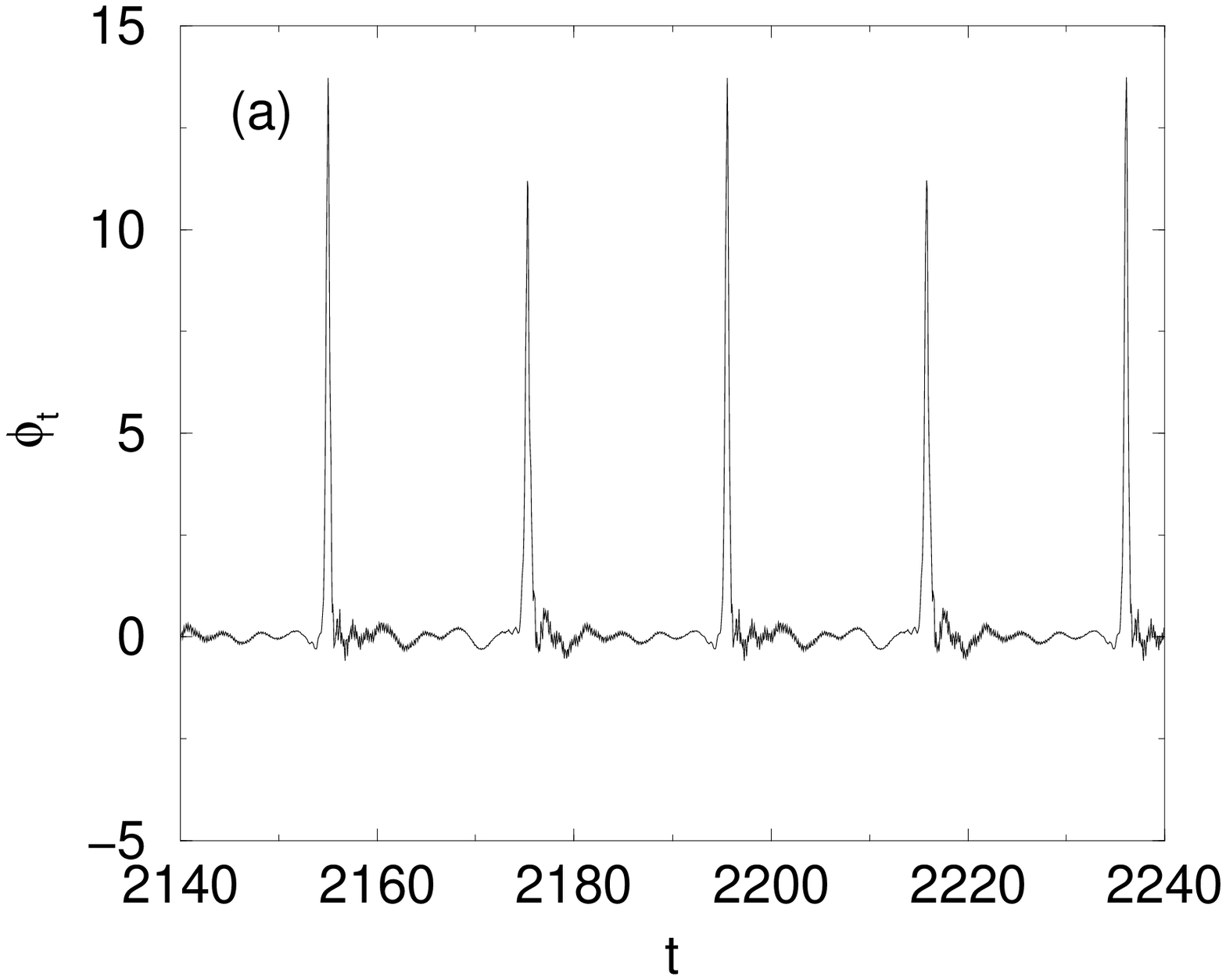,width=8.5cm,angle=0}}
\centerline{\psfig{figure=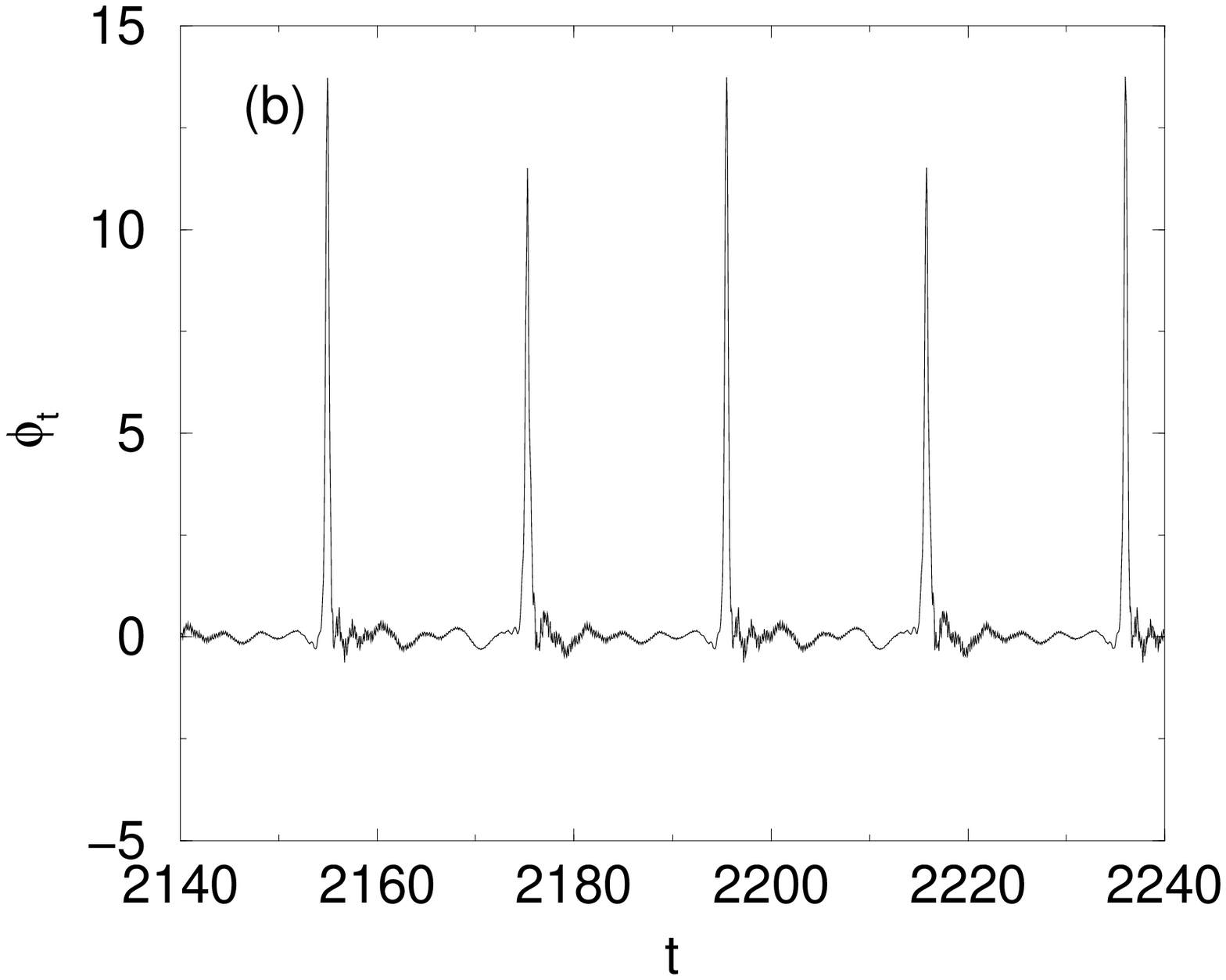,width=8.5cm,angle=0}}
\caption{Instantaneous voltage in the middle of the junction $(x=0)$
vs time $t$, for the solutions in the first ZFS.
$l=20$, $\gamma=0.01$, $I=1.6$: (a) $ff$,
(b) $faf$. The pairing state is $B_{1g}$-wave.}
\label{ft1ZFS.fig}
\end{figure}

The different character of the various fluxon solutions
can also be seen from the plot of the instantaneous
voltage $\phi_t$ at the center of the junction
for the various fluxon configurations.
This plot is seen in Fig. \ref{ft1ZFS.fig} for the solutions
regarding the first ZFS. During
the time of one period three peaks appear in this plot by the time when
the fluxon (antifluxon) passes through the junction center.
For the first ZFS the $\phi_t$ vs $t$ plot
can be used to
probe the existence of $ff$ or $faf$ at the junction center.
The height of the middle peak is smaller
for a bound $ff$ at the junction center, than for a bound $faf$.
The plot of $\phi_t$ at the edges shows two peaks
at time instants which differ by half
a period. Note that the characteristic oscillations
of $\phi_t$ between the peaks are due to the oscillation
of the bound solution about the junction center. These oscillations
have the same amplitude for the $ff$ and $faf$ cases.

For the $B_{1g}$ case the bound fluxon and antifluxon have equal 
magnitudes or contain equal magnetic flux 
and the critical current is the same as seen in Fig. \ref{IV.fig}(a). 
However for the $E_u$ case the $ff$ contains less 
flux than $faf$ and has smaller critical current as seen in 
Fig. \ref{IV.fig}(b). 
For the $B_{1g}\times E_u$ case the $ff$ contains more
flux than $faf$ and has greater critical current as seen in 
Fig. \ref{IV.fig}(c). 

For the second ZFS
multiple 
solutions exist in which two fluxons are propagating in the junction 
in different configurations. These solutions 
can be classified depending on the fluxon separation as seen in 
Fig. \ref{IV.fig} and give distinct 
critical currents in the $I-V$ diagram. 
For all the modes in the second ZFS we can estimate the value of the
constant dc-voltage
where they occur as follows.
A full period of motion back and forth takes time $T=40$, and
since the overall phase advance is $8\pi$, in the relativistic limit
where $u=1$ reached at high currents, the dc voltage across the
junction will be $V=0.628$.
So compared to the case of conventional $s$-wave superconductors
junction we observe several curves for the second ZFS
depending on the relative distance between the fluxons
and this may be used to probe the presence of intrinsic magnetic
flux.

\begin{figure}
\centerline{\psfig{figure=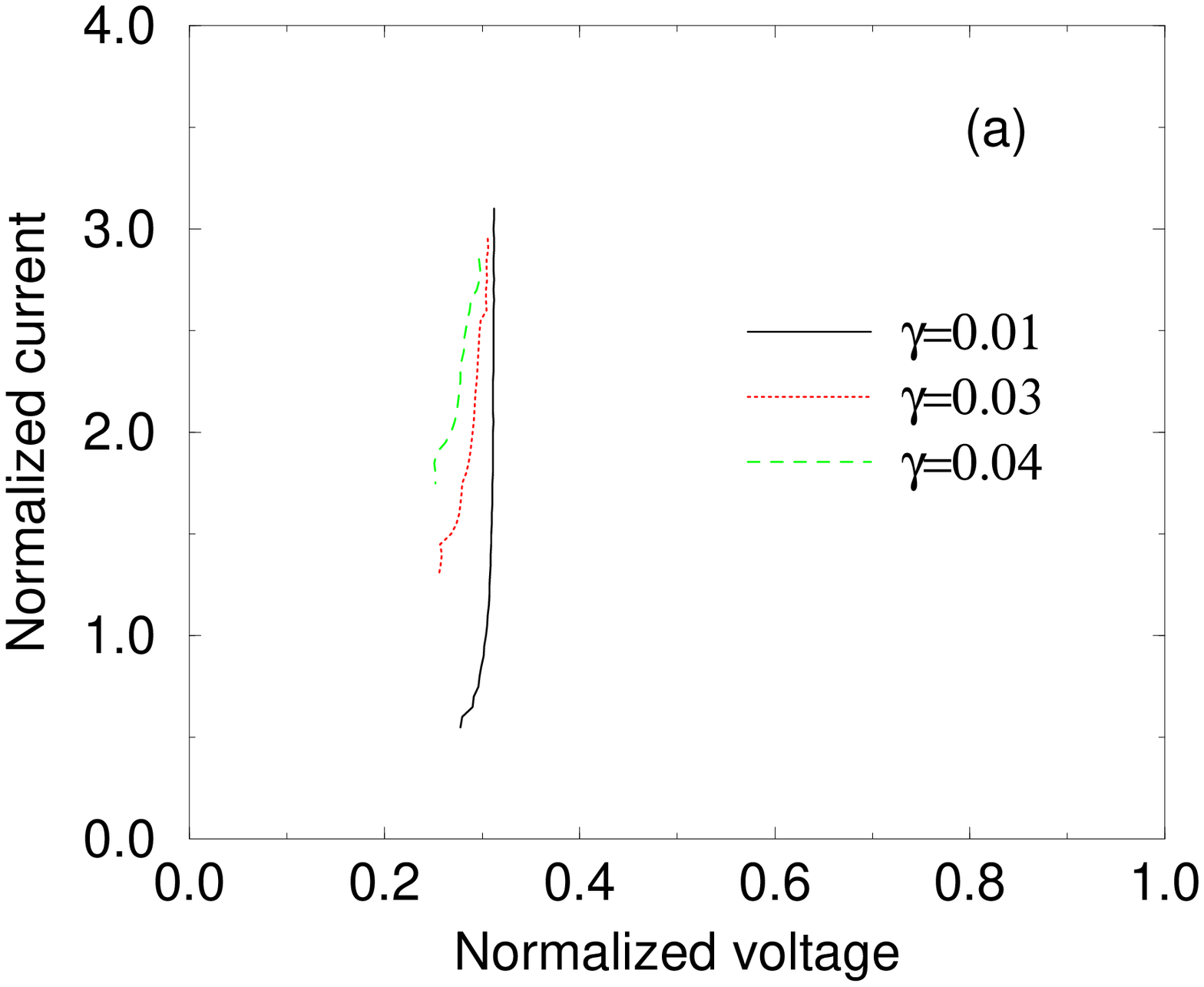,width=8.5cm,angle=0}}
\centerline{\psfig{figure=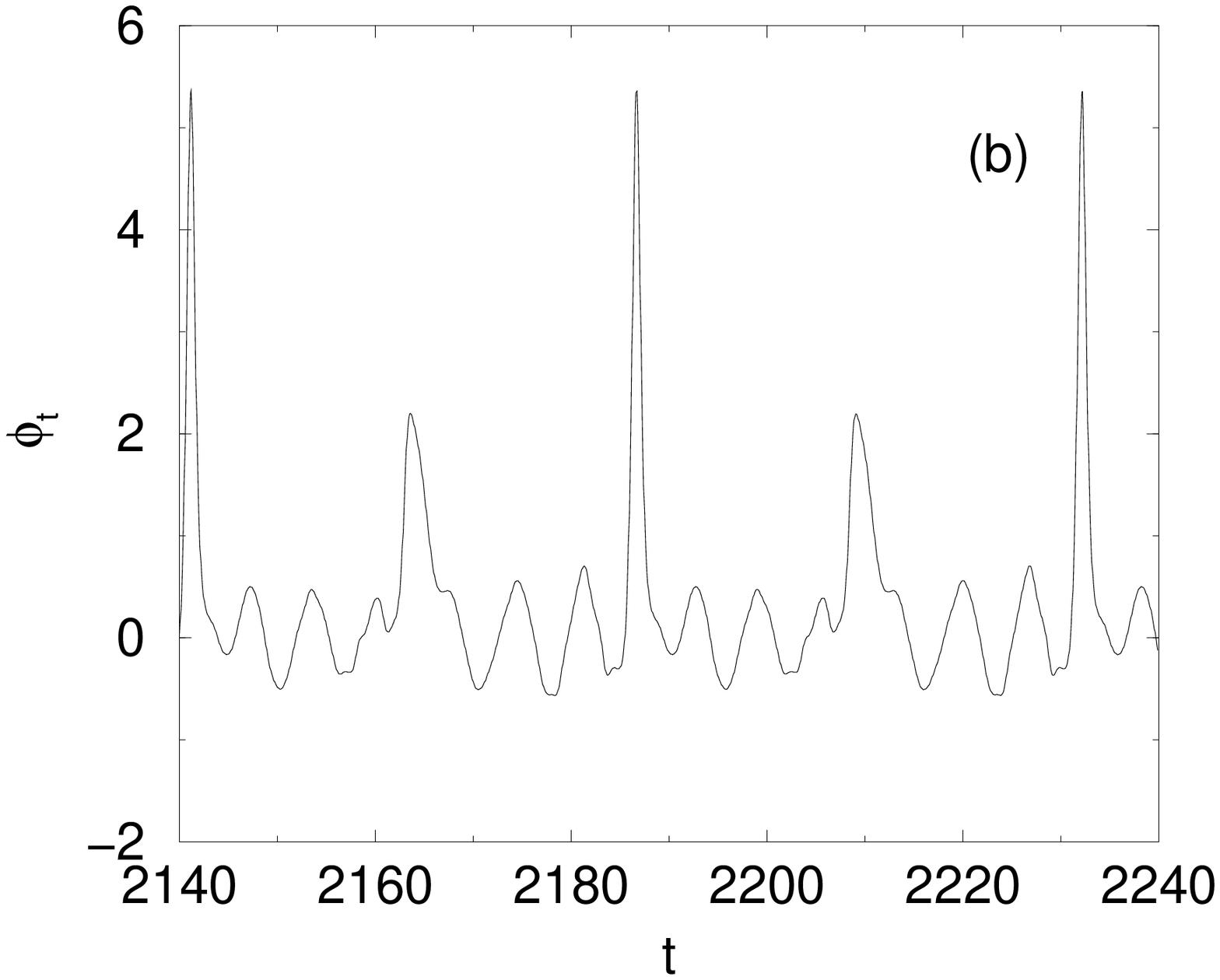,width=8.5cm,angle=0}}
\caption{
(a) Normalized current versus the normalized voltage
for the inline geometry, for the $ff$ solution, for different values 
of the damping constant $\gamma$. $l=20$. The pairing state is $B_{1g}$-wave.
(b) Instantaneous voltage in the middle of the junction $(x=0)$
vs time $t$, for the solution $ff$,
$l=20$, $I=1.6$.
The pairing state is $B_{1g}$-wave. The damping constant is $\gamma=0.03$.}
\label{damping.fig}
\end{figure}

We also considered damping effects due to the quasiparticles
because the Josephson junctions made of high-Tc materials are 
highly damped. 
In the $I-V$ curves higher damping shifts the $I-V$ curve
upwards (see Fig. \ref{damping.fig}(a)) 
and the fluxon reaches the critical current velocity 
only for currents that are very close to the 
critical current where the jump to the resistive branch occurs. 
In this case the ZFS are not 'vertical'. 
In the plot of the instantaneous voltage at the middle of the junction 
versus the time (Fig. \ref{damping.fig}(b)) 
the difference in height between the peaks 
when the $F$ or the $AF$ interacts with the bound $ff$ is larger 
for greater values of damping. The oscillations between the peaks 
become very large by increasing the damping and finally make the 
solution unstable.
We believe that these solutions are more stable for small values 
of the damping at least for inline boundary conditions. 
In the overlap case (not presented in the figure) these solutions 
are more stable compared to the inline case.
 
\section{short length limit}
\begin{figure}
\centerline{\psfig{figure=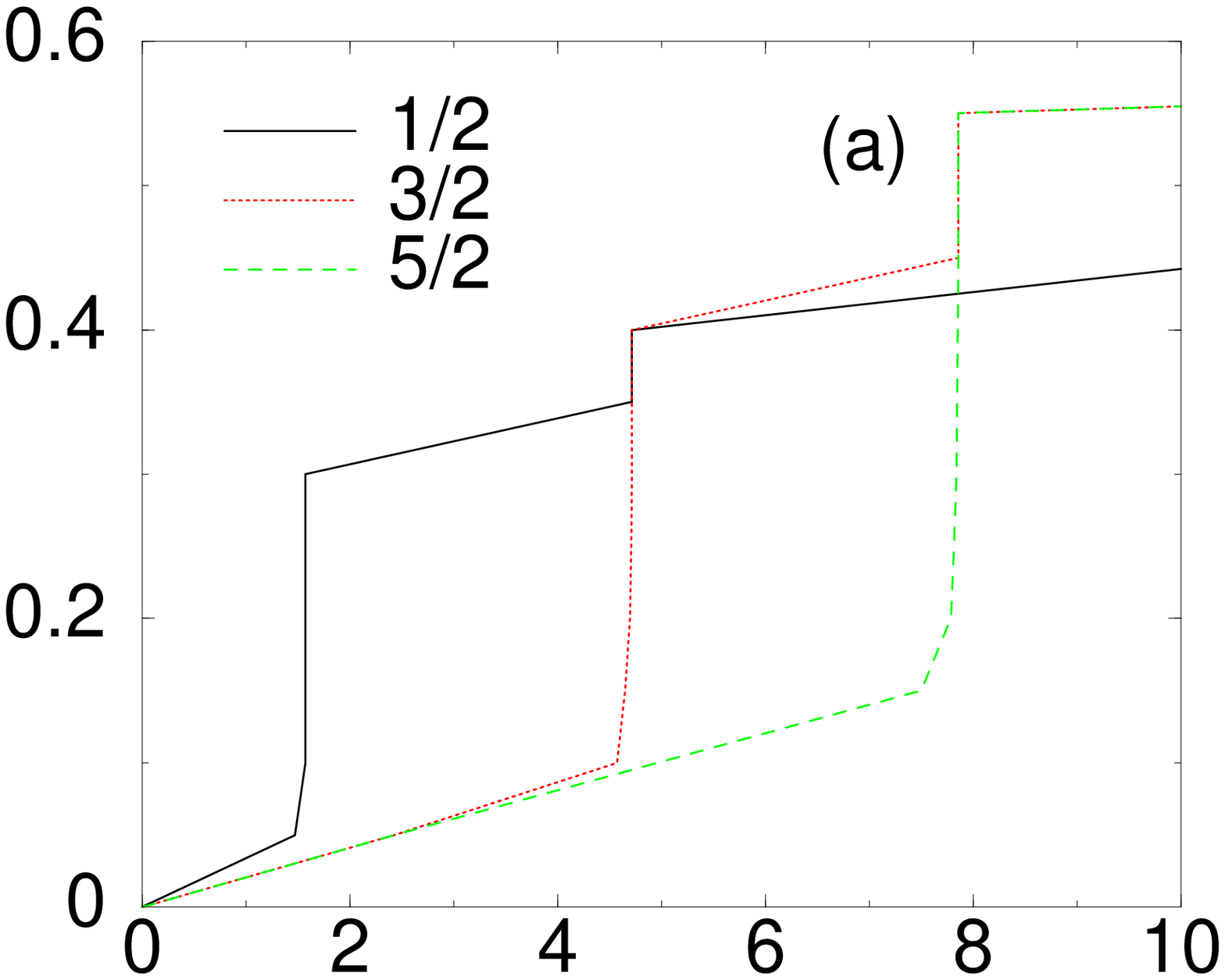,width=5.5cm,angle=0}}
\centerline{\psfig{figure=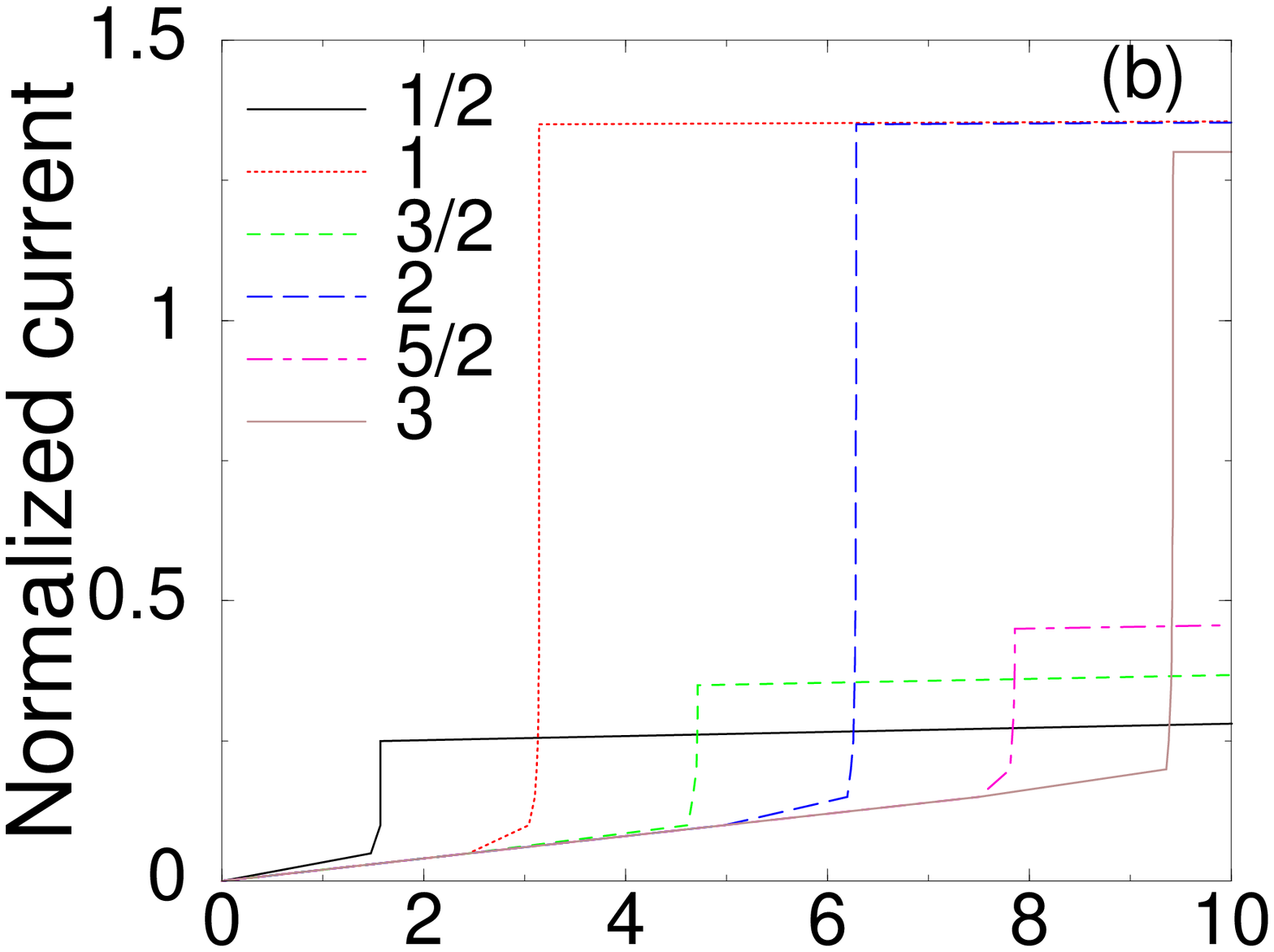,width=5.5cm,angle=0}}
\centerline{\psfig{figure=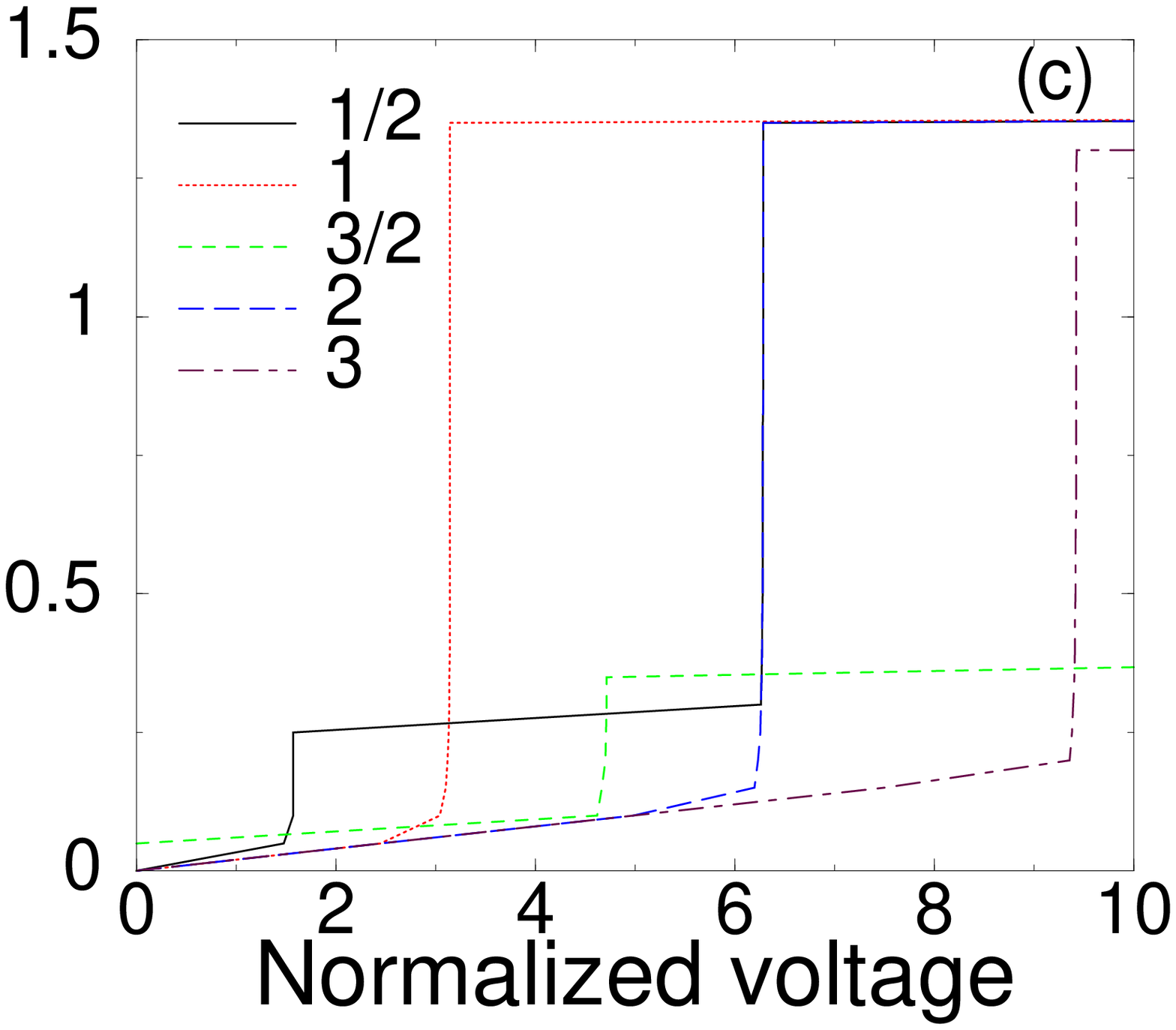,width=5.5cm,angle=0}}
\caption{Normalized current versus the normalized voltage
for the inline geometry, 
for a junction of length $l=2$, $\gamma=0.01$. The different modes are labeled 
by the value of the normalized voltage divided by $\pi$ in which 
they occur.
The pairing state is (a) $B_{1g}$-wave, (b) $E_u$-wave, 
(c) $B_{1g}\times E_u$-wave.
}
\label{IV2.fig}
\end{figure}

When the junction length is small $l=2$ the fractional fluxon
or antifluxon
does not remain confident at the junction's center but is able to 
move along the junction as a bunched type solution. 
The moving fluxon configuration could have fractional flux and additional
steps are introduced in the $I-V$ diagram. 
We plot in Fig. \ref{IV2.fig} the $I-V$ characteristics for the 
$B_{1g}$, $E_u$, $B_{1g}\times E_u$-wave pairing states, for the small junction length $l=2$. 

\begin{figure}
\centerline{\psfig{figure=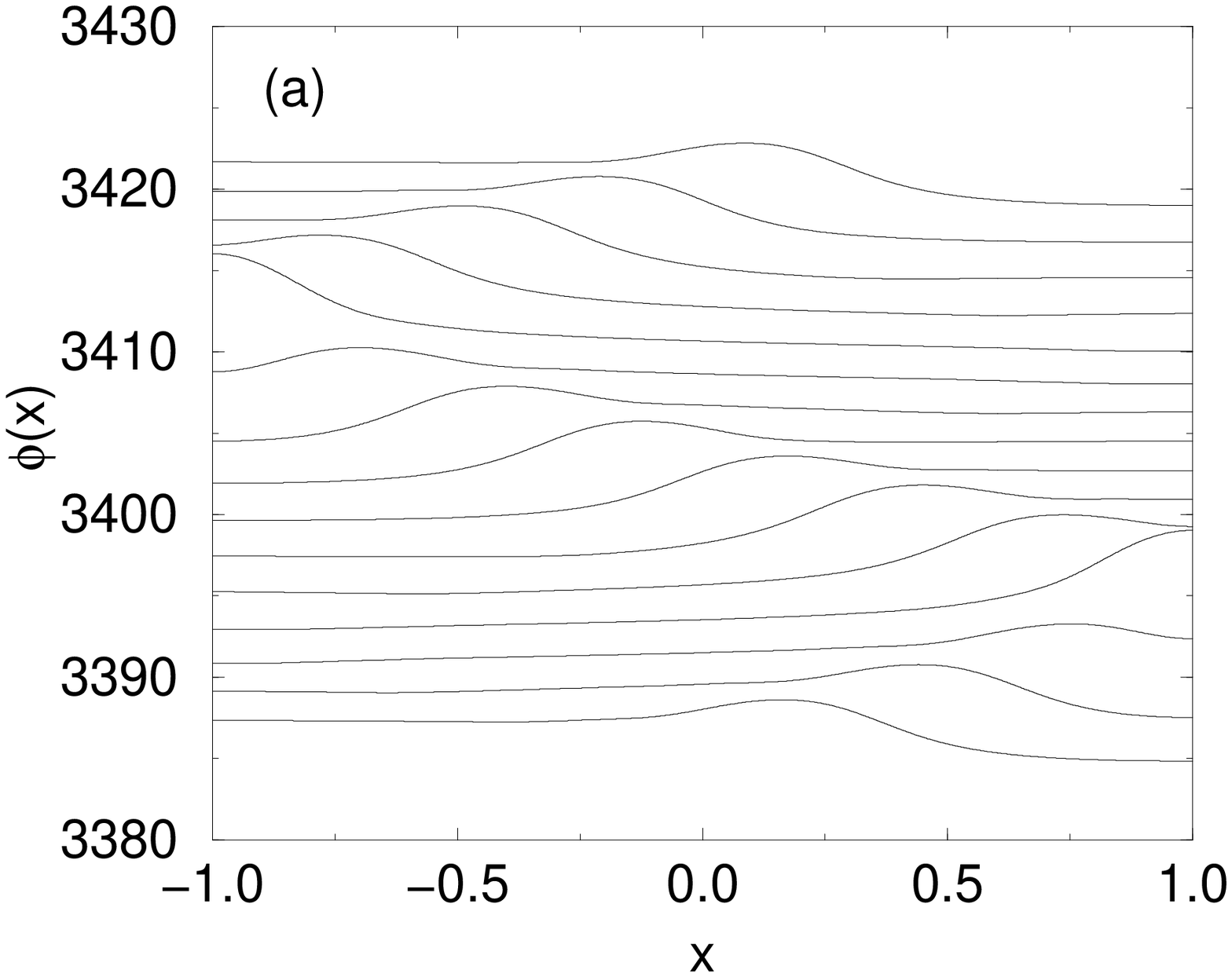,width=8.5cm,angle=0}}
\centerline{\psfig{figure=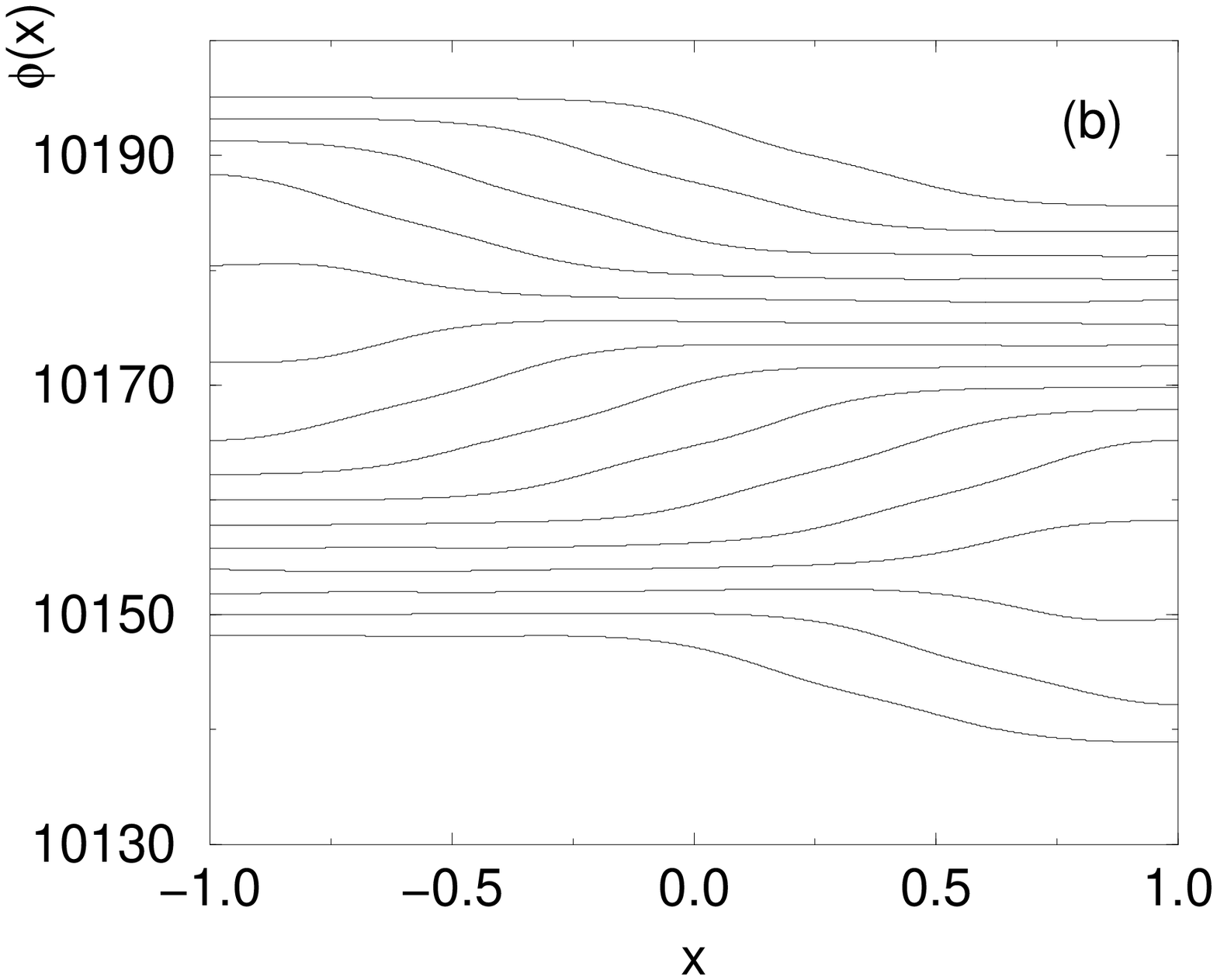,width=8.5cm,angle=0}}
\centerline{\psfig{figure=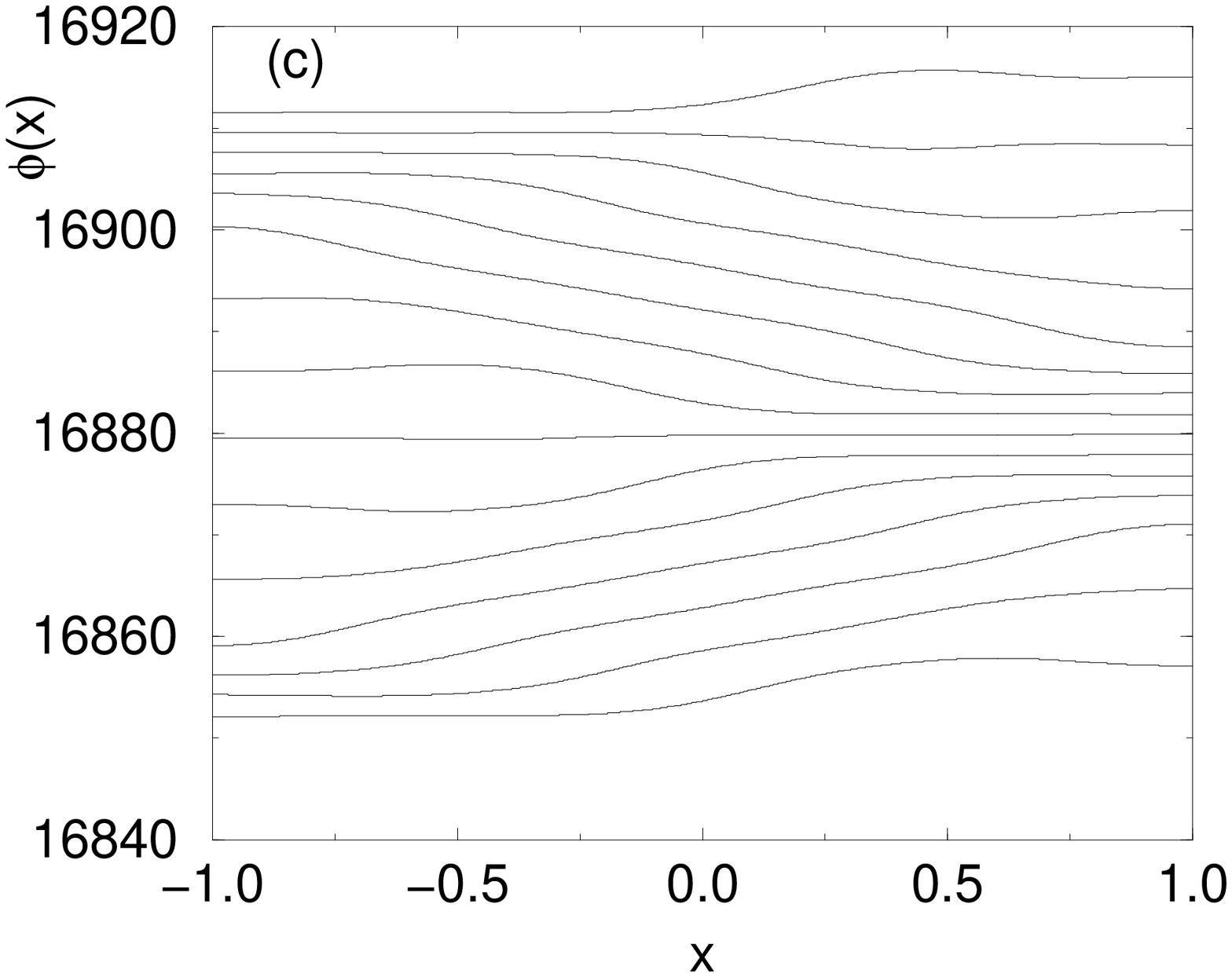,width=8.5cm,angle=0}}
\caption{Phase $\phi(x)$ vs $x$
for the solutions in the first three ZFS,
at various instants, during one period
separated by $\Delta \tau=0.2$.
The pairing state is $B_{1g}$-wave.
The curves are shifted by $0.5$ to avoid overlapping.
$l=2$, $I=0.25$, $\gamma=0.01$.
}
\label{l2.fig}
\end{figure}

\begin{figure}
\centerline{\psfig{figure=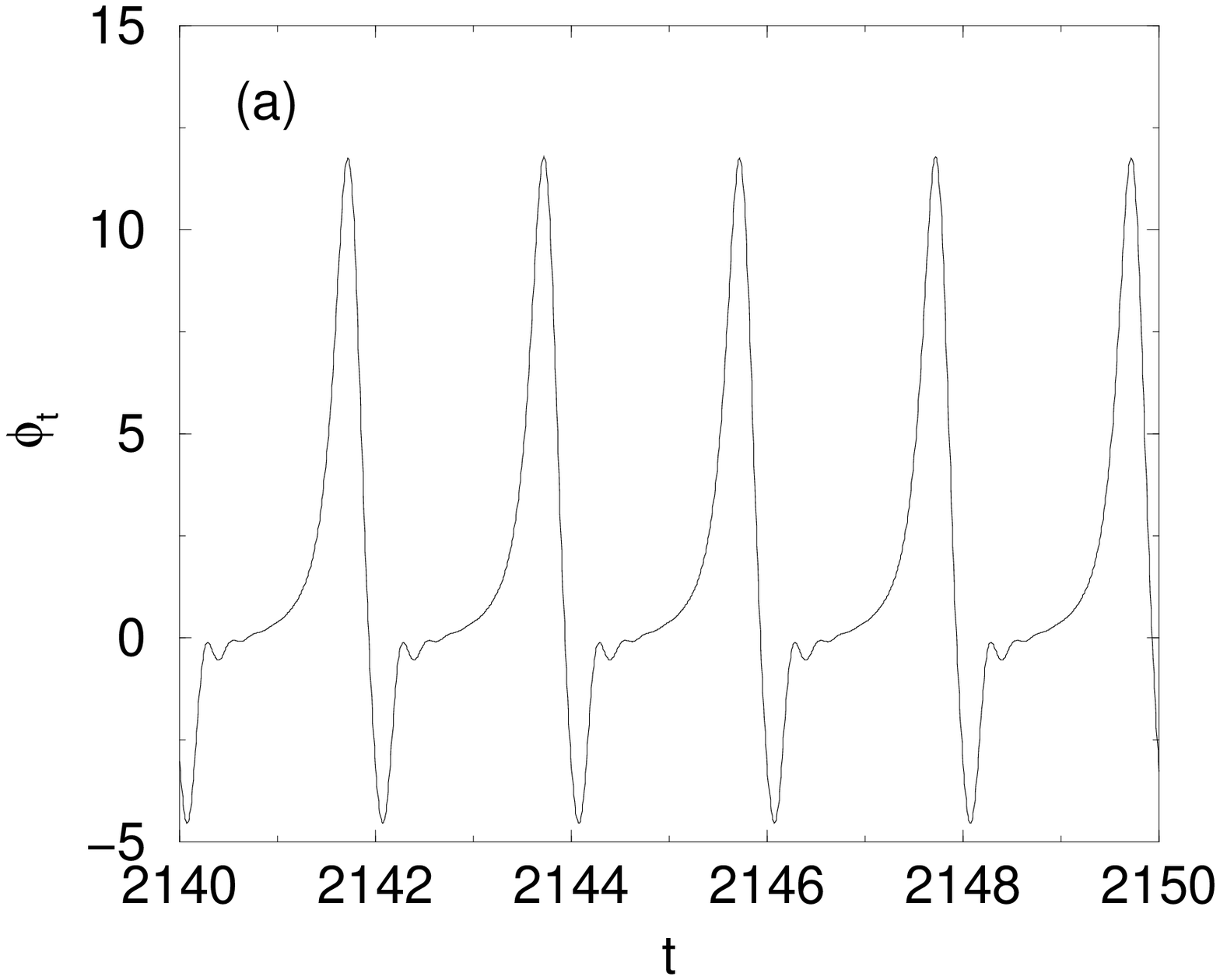,width=8.5cm,angle=0}}
\centerline{\psfig{figure=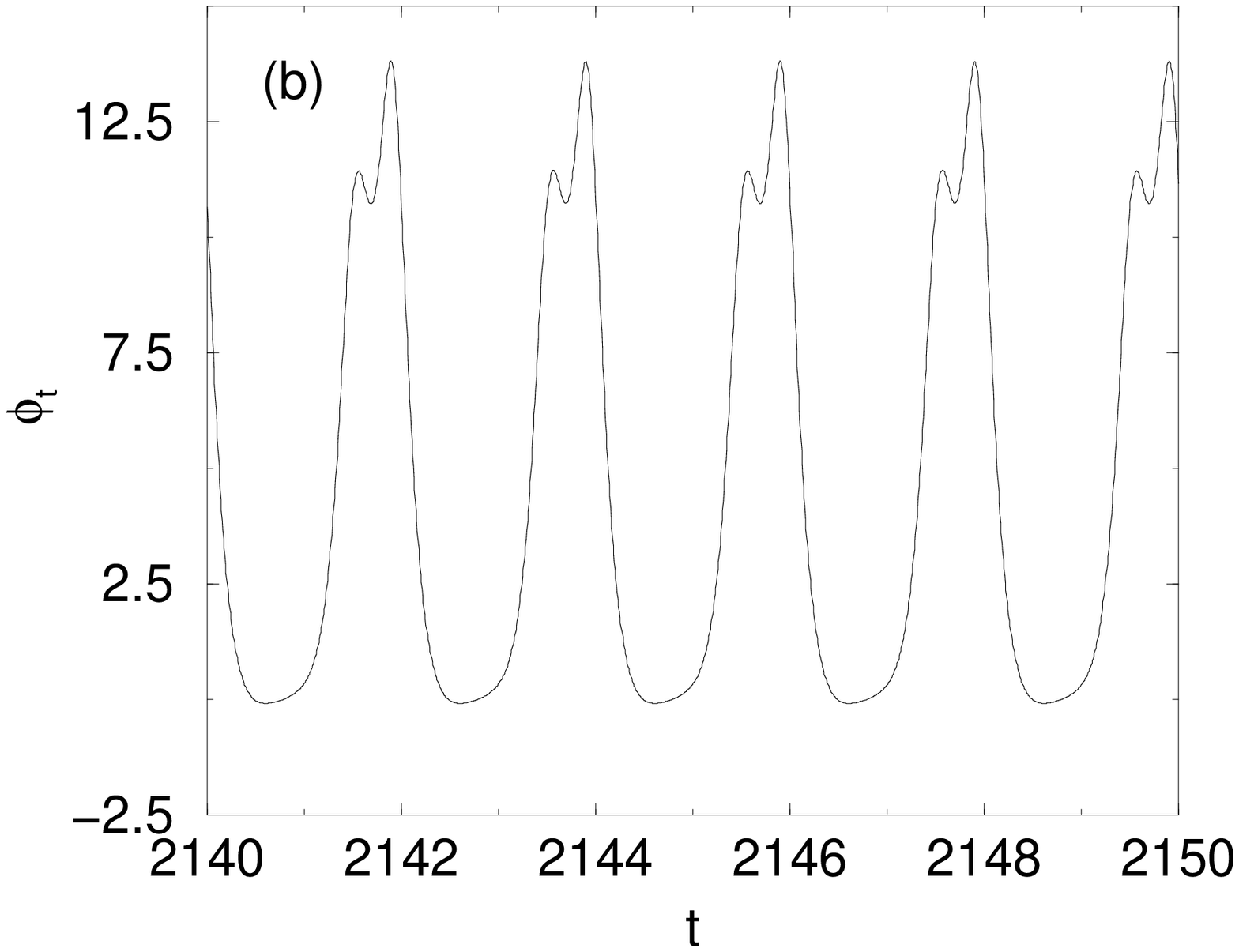,width=8.5cm,angle=0}}
\centerline{\psfig{figure=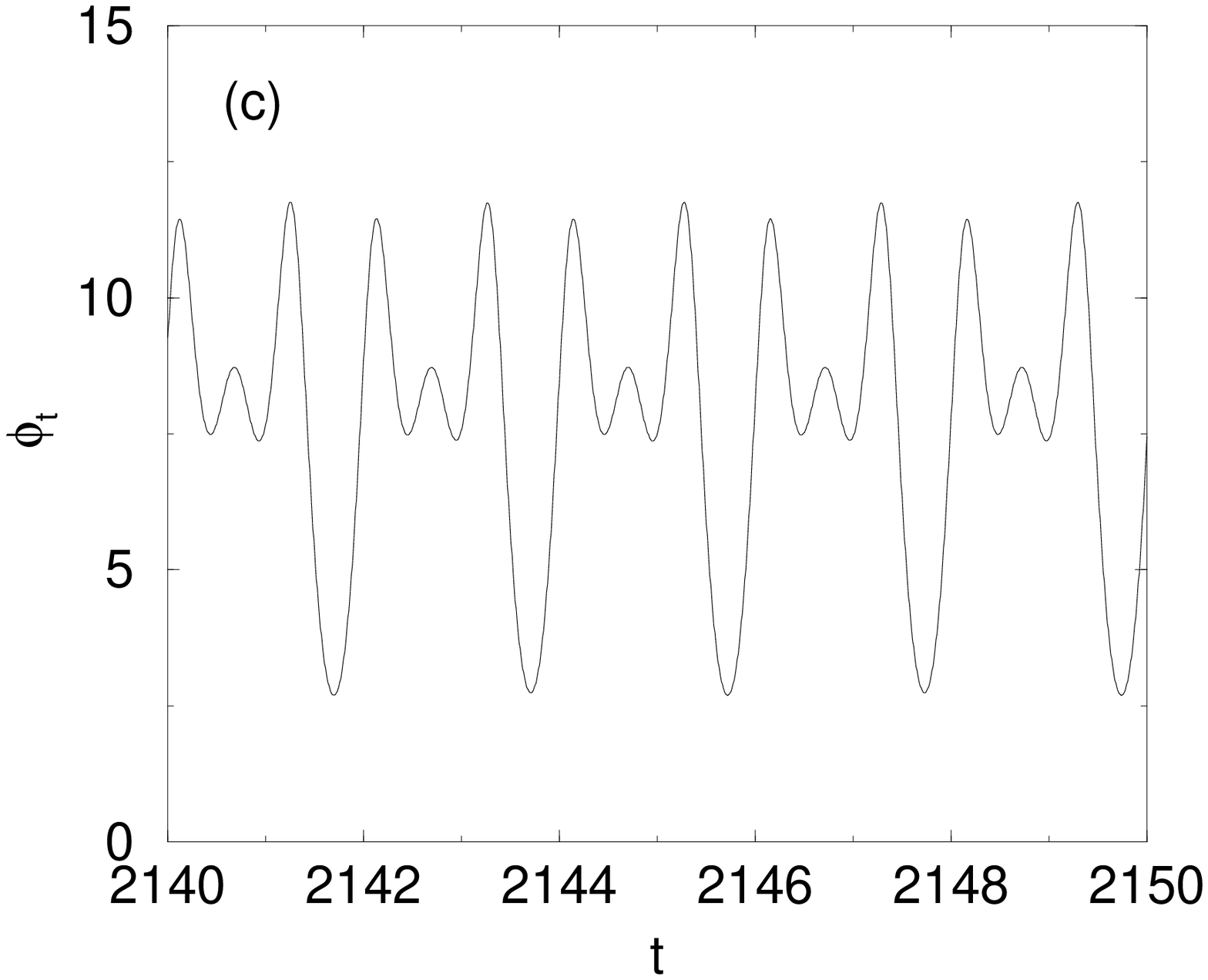,width=8.5cm,angle=0}}
\caption{Instantaneous voltage in the middle of the junction $(x=0)$
vs time $t$, for the solutions in the first three ZFS.
The pairing state is $B_{1g}$-wave.
$l=2$, $\gamma=0.01$, $I=0.25$.
}
\label{ftl2.fig}
\end{figure}

For the $B_{1g}$ pairing state, and the first ZFS 
the moving fluxon configuration to the right is a 
combination of a fractional fluxon and a fractional 
antifluxon which contains half-integer magnetic flux 
(see Fig. \ref{l2.fig}(a)). 
When it hits the right boundary it transforms to a configuration with opposite 
sign of flux which moves to the left. 
Then it hits the left boundary and transforms into a configuration that moves 
to the center and
the period is completed.
In the relativistic limit reached at high currents a full period
of motion back and forth takes time $t=2l/c_S=4$ and since the overall phase
advance is $2\pi$ the normalized voltage will be $V=\frac{\pi}{2}$ as seen in 
Fig. \ref{IV2.fig}(a) for the solution labeled as $1/2$.
Note that this value
is half than the case where a full fluxon moves into the junction.
In Fig. \ref{ftl2.fig}(a) we plot the $\phi_t$ vs $t$ at the center of
the junction
where the successive peaks correspond to the passage of the fluxon
combination from the junction center.
The $\phi_t$ pulse is composed of two peaks, a positive and
a negative one, corresponding to the $AF$ and the $ff$ for the fluxon traveling
in the forward direction.
The pulse structures corresponding to the forward
and backward directions are the same due to the symmetrical configuration
of the fluxons traveling in the forward and backward directions.

It is also possible to have solutions where
an integer fluxon plus a fractional fluxon is propagating into the junction
(see Fig. \ref{l2.fig}(b)).
In this case the magnetic flux is equal to $1.5$. In a junction of length $l=2$
the propagating fluxon accomplishes an overall phase advance
of $6\pi$ in a full period $T=4$.
Thus the voltage across the
junction will be $V=3\pi/2$ as seen in
Fig. \ref{IV2.fig}(a) for the solution labeled as $3/2$. 
In Fig. \ref{ftl2.fig}(b) we plot the $\phi_t$ vs $t$ at the center of
the junction
where the successive peaks correspond to the passage of the fluxon
combination from the junction center.
The moving fluxon or antifluxon has internal structure and therefore
a double peak structure appears in the $\phi_t$ vs $t$ diagram.

Finally in Fig. \ref{l2.fig}(c) we present $\phi(x)$, for the case where
two integer fluxons with a fractional fluxon in between
move into the junction.
This configuration contains magnetic flux equal to $2.5$.
Thus the dc voltage across the
junction will be $V=5\pi/2$ as seen in Fig. \ref{IV2.fig}(a) 
for the solution labeled as $5/2$.
In Fig. \ref{ftl2.fig}(c) we plot the $\phi_t$ vs $t$ at the center of
the junction
where the periodic pattern of three peaks corresponds to the passage
of the integer fluxons and the half integer fluxon from the
junction's center.

So for the $B_{1g}$ 
pairing state the ZFS appear at values of the normalized voltage 
that are displaced by $1/2$ compared to the case of $s$-wave superconductor
junction. This value of the voltage corresponds to the intrinsic
phase shift. This is analogous to the shift of the critical current 
versus the magnetic flux in a corner squid 
of $d$-wave superconductors \cite{vanh}
and is 
expected to be confirmed by experiment.

For the $E_u$ and $B_{1g}\times E_u$ cases, 
we can have additional solutions where the moving fluxon has integer 
flux and the voltage steps appear at values $V=n\pi$ in addition 
to $V=\frac{n\pi}{2}$ as seen in
Figs. \ref{IV2.fig}(b) and \ref{IV2.fig}(c) for the solutions labeled 
with integer numbers.
For the $B_{1g}$-wave case the forward and backward configurations 
are symmetric and successive peak structures have the same form. 
In the $E_u$ and $B_{1g}\times E_u$ cases, successive peaks, 
corresponding to the 
structure of the fluxon configuration that moves in the forward and backward 
direction,
have 
different amplitudes indicating that the fluxon configurations moving in the 
forward and backward directions have different structure. 

We examined also the case where the pairing symmetry of the 
superconductor is $d+is$. 
In this case due to the difference in the flux content of the static
solutions \cite{stefan1}, 
the critical currents for the $ff, faf$ modes of the first ZFS
do not coincide.

\section{qubit implementation}
Also the frustrated junction could be considered as a way to build 
a qubit.
This idea has also been implemented using 
$s$-wave/$d$-wave/$s$-wave junction 
exhibiting a degenerate ground state and a double-periodic 
current-phase relation \cite{ioffe}, or superconducting Josephson 
junction arrays \cite{averin}. 
Also the dynamics of a Josephson charge qubit, coupled capacitively 
to a current biased Josephson junction has been studied \cite{hekking}.
The two segments of the frustrated junction have characteristic 
energies $E(\phi)$, and $E(\phi+\pi)$ and the resulting energy has 
minima at $\pm \pi/2$.
The system exhibits a
degenerate ground state.
The bound $ff$, $faf$ can be considered as the two quantum levels of 
our system. By applying an external current a bunched type solution 
containing $0.5(-0.5)$ flux is generated 
if the actual ground state of the system is $ff(faf)$ which propagates 
into the junction to the left(right) and generates a voltage pulse 
which can be determined at the boundaries. 
A possible method to generate the desired ground state for our system,
i.e., $ff$, or $faf$ is to apply an external magnetic field
(positive or negative) and then to slowly decrease it. Depending on the
sign of the external field the system will go either in the $ff$ or
in the $faf$ state.

In the $E_u$, $B_{1g}\times E_u$ 
cases the actual ground state of the system is not degenerate, i.e., the $ff$ 
and $faf$ carry different flux. Moreover the traveling fluxon caring half 
the flux quantum to the left 
has different structure than the one traveling to the right and  
the $\phi_t$ at the ends are different for the $ff$, $faf$.
In the long length limit the presence of the $ff$, $faf$ at the center can 
be deduced from a measurement of the $\phi_t$ at the edges since the 
direction of the moving integer flux depends on the sign of the intrinsic flux. 
In this case the intrinsic flux can not escape from the junction's center.

\section{conclusions}
The fluxon dynamics in frustrated
Josephson junction with $p$, $d$, and $f$-wave pairing symmetry,
is different in the long and short junction limit.
When $l$ is large, the bound intrinsic flux remains confined at $x=0$, 
and the moving integer fluxon or antifluxon interacts with it only when 
it approaches the center.
However when the length is small the bound fluxon becomes able to move as 
a bunched type solution. For $d$-wave junction the $I-V$ curves are 
displaced by a voltage that corresponds to the intrinsic phase shift. 

The resonant fluxon motion also can be determined experimentally 
in one-dimensional ferromagnetic $0-\pi$ junctions 
where the width of the ferromagnetic oxide layer determines 
the region of the junction where the Josephson critical current 
is positive or negative \cite{bourgeois,kontos}. In this case 
the junction contour does not have to change as in the case of a 
corner junction. However the change of the junction contour was 
not a problem for the realization of the intrinsic fluxon in 
the static problem, so way should it prevent the realization of its motion?

A final comment is that the frustrated junctions that we consider in this 
paper are realized in the ($ab$)-plane of the unconventional 
superconductors due to the sign change of the order parameter.
This type of junctions is different from the series array of intrinsic 
Josephson junctions in high-Tc supercondcutors where the 
Josephson effect is observed in the $c$-axis, for instance in 
Bi$_2$Sr$_2$CaCu$_2$O$_8$ crystals \cite{kleiner}.  
 
\section{acknowledgements}
Part of this work was done at the 
Department of Physics, University of Crete, Greece. The author wishes 
to thank Dr. N. Lazarides for valuable discussions.

\begin{table}
\caption{
We present the characteristic phases $\phi_{c1}, \phi_{c2}$ for the various 
pairing symmetries.
$\phi_{c1}, \phi_{c2}$
is the extra phase difference in the two edges of the corner junction due to the
different orientations, of the $a$-axis of the crystal lattice. 
}
\begin{tabular}{ccc}
	Pairing state & $\phi_{c1}$ & $\phi_{c2}$\\ \hline
	$B_{1g}$ & $0$ & $\pi$ \\
	$E_u$ & $0$ & $-\pi/2$\\
	$B_{1g}\times E_u$ & $0$ & $\pi/2$
\end{tabular}
\label{tablephic}
\end{table}

\end{document}